\documentclass[12pt,letterpaper]{article}

\usepackage{amsmath}

\usepackage{textcomp}
\usepackage[disable]{todonotes}
\usepackage{listings}

\usepackage{algorithm}
\usepackage{algpseudocode}
\usepackage{pifont}

\newcommand{\NAME}{\textit{HPVM}\/}

 \newcommand{\comment}[1]{}  %comment not showed
\lstdefinelanguage{llvm}{
  morecomment = [l]{;},
  morestring=[b]",
  sensitive = true,
  classoffset=0,
  morekeywords={
    declare, global, constant,
    internal, external, private,
    linkonce, linkonce_odr, weak, weak_odr, appending,
    common, extern_weak,
    thread_local, dllimport, dllexport,
    hidden, protected, default,
    except, deplibs,
    volatile, fastcc, coldcc, cc, ccc,
    x86_stdcallcc, x86_fastcallcc,
    ptx_kernel, ptx_device,
    signext, zeroext, inreg, sret, nounwind, noreturn,
    nocapture, byval, nest, readnone, readonly, noalias, uwtable,
    inlinehint, noinline, alwaysinline, optsize, ssp, sspreq,
    noredzone, noimplicitfloat, naked, alignstack,
    module, asm, align, tail, to,
    addrspace, section, alias, sideeffect, c, gc,
    target, datalayout, triple,
    blockaddress,
  },
  classoffset=1, keywordstyle=\color{purple},
  morekeywords={
    fadd, sub, fsub, mul, fmul,
    sdiv, udiv, fdiv, srem, urem, frem,
    add, sub,
    and, or, xor,
    icmp, fcmp,
    eq, ne, ugt, uge, ult, ule, sgt, sge, slt, sle,
    oeq, ogt, oge, olt, ole, one, ord, ueq, ugt, uge,
    ult, ule, une, uno,
    nuw, nsw, exact, inbounds,
    phi,
    call,
    select, shl, lshr, ashr, va_arg,
    trunc, zext, sext,
    fptrunc, fpext, fptoui, fptosi, uitofp, sitofp,
    ptrtoint, inttoptr, bitcast,
    ret,
    define,
    br, indirectbr, switch, invoke, unwind, unreachable,
    malloc, alloca, free, load, store, getelementptr,
    extractelement, insertelement, shufflevector,
    extractvalue, insertvalue,
    type
  },
  alsoletter={\%,.},
  keywordsprefix={@},
}

\usepackage{titling}
\newcommand{\subtitle}[1]{%
  \posttitle{%
    \par\end{center}
    \begin{center}\Large#1\end{center}
    \vskip0.5em}%
}

\begin{document}

\title{\NAME{}: A Portable Virtual Instruction Set for Heterogeneous Parallel Systems}
\subtitle{Technical Report}
\date{\today}

\author{Prakalp Srivastava\\
        University of Illinois at Urbana Champaign\\
        psrivas2@illinois.edu
        \and
        Maria Kotsifakou\\
        University of Illinois at Urbana Champaign\\
        kotsifa2@illinois.edu
        \and
        Vikram Adve\\
        University of Illinois at Urbana Champaign\\
        vadve@illinois.edu
} % Author name

\maketitle

%------------------------------------------------------------------------------
\section{Introduction}
\label{sec:introduction}
%------------------------------------------------------------------------------

We describe a programming abstraction for heterogeneous parallel hardware,
designed to capture a wide range of popular parallel 
hardware, including GPUs, vector instruction sets and multicore CPUs.
Our abstraction, which we call \NAME{}, is
\emph{a hierarchical dataflow graph with shared memory and
vector instructions}.
We use \NAME{} to define both a virtual instruction set (ISA) and also
a compiler intermediate representation (IR).
The virtual ISA aims to 
achieve both functional portability and performance portability across
heterogeneous systems,
while the compiler IR aims to enable effective code 
generation and optimization for such systems.

\NAME{} effectively supports \emph{all} forms of parallelism used to achieve 
computational speedups (as opposed to concurrency),
including task parallelism, coarse-grain data parallelism, fine-grain data 
parallelism, and pipelined parallelism.
\NAME{} also enables flexible scheduling and tiling:
different nodes in the dataflow graph can be mapped flexibly to different
combinations of compute units, and the graph hierarchy expresses memory
tiling, essential for achieving high performance on GPU and CPU targets.

%------------------------------------------------------------------------------
%\section{Motivation}
%\label{sec:motivation}
%------------------------------------------------------------------------------

%------------------------------------------------------------------------------
\section{Design}
\label{sec:design}
%------------------------------------------------------------------------------

In this section, we describe the design of the \NAME{} parallel abstractions,
and how they are defined as an extension to the LLVM Internal Representation
(IR).  We also discuss briefly some features of the design that are primarily
for performance rather than correctness (or ```functionality'').

Section~\ref{sec:example:hpvm} contains a sample code written in \NAME{}.
Throughout this section, we will refer to Section~\ref{sec:example:hpvm} and use
the example to illustrate certain parts of the \NAME{} design.

%------------------------------------------------------------------------------
\subsection{Dataflow Graph}
\label{sec:design:dfgraph}
%------------------------------------------------------------------------------
In \NAME{}, a program is represented as a hierarchical dataflow
graph (DFG) with side effects, where 
nodes represent units of execution, and
edges between nodes describe the explicit data transfer requirements.
Each dataflow node in a DFG can either be a 
\emph{leaf node} or an \emph{internal node}.
An internal node contains a complete dataflow graph (child graph), and the child
graph itself can have internal nodes and leaf nodes.
A leaf node contains a mixture of scalar and vector code,
expressing actual computations.
Figure~\ref{fig:sgemmdfg} shows the dataflow graph of an application, sgemm,
based on \NAME{} abstractions. Dataflow nodes {\tt SgemmLeaf} and
{\tt Allocation} are leaf nodes, while {\tt SgemmInternal} and {\tt SgemmRoot}
are internal nodes containing other dataflow graphs.

Dataflow edges can be ordinary edges, denoting a one-time data transfer, or
streaming edges denoting that data items will be repeatedly transferred through
this edge, and thus will need to be processed by repeated executions of the sink
dataflow node. Figure~\ref{fig:sgemmdfg} shows two (ordinary) dataflow edges
from {\tt Allocation} node to {\tt SgemmLeaf} node, representing transferring of
two data items.

To express data parallelism, which is inherently the replication of computation
over data, we allow a single static dataflow node to represent multiple
dynamic instances of the node. The dynamic instances of a
node are required to be independent of each other, i.e., can be executed
in parallel. For example, in sgemm (subsection~\ref{sec:example:hpvm}),
{\tt SgemmLeaf} node has multiple dynamic instances, each computing different
output elements of the multiplication result. Note that 
figure~\ref{fig:sgemmdfg} shows the static dataflow graph, thus the replication
of {\tt SgemmLeaf} is not depicted.

Consequently, dataflow edges are also replicated.
We provide two replication mechanisms:
\begin{itemize}
\item ``all-to-all'': all dynamic instances of the source node are connected
      with all dynamic instances of the sink node, thus expressing a
      synchronization barrier between the two groups of nodes
\item ``one-to-one'': each dynamic instance of the source dataflow node is
      connected with a single corresponding instance of the sink node.
      One-to-one replication requires that the grid structure (number of
      dimensions and the extents in each dimension) of the source and sink
      nodes be identical.
\end{itemize}

%------------------------------------------------------------------------------
\subsection{HPVM Extension to LLVM}
\label{sec:design:intrinsics}
%-----------------------------------------------------------------------------

%------------------------------------------------------------------------------
\begin{table*}[ht]
{\footnotesize
\begin{tabular}{p{0.5\textwidth}|p{0.5\textwidth}}
\hline
\multicolumn{2}{l}{\em Intrinsics for Constructing Graphs}\\
\hline
{\tt\footnotesize i8* {\bf llvm.hpvm.createNode1D}(Function* F, i32 n)}	&
	{\small Create node with $n$ dynamic instances executing node function {\tt F} 
	(similarly {\bf llvm.hpvm.createNode2D/3D})}\\
{\tt\footnotesize void {\bf llvm.hpvm.createEdge}(i8* Src, i8* Dst,
			\hspace*{2em}i32 sp, i32 dp, i1 ReplType, i1 Stream)}	&
	{\vspace*{0.1\baselineskip}
	 \small Create edge from output $sp$ of node {\tt Src} to input $dp$ of node {\tt Dst}}\\
{\tt\footnotesize void {\bf llvm.hpvm.bind.input}(i8* N, i32 ip, 
			\hspace*{3em}i32 ic, i1 Stream)}	&
	{\small Bind input $ip$ of current node to input $ic$ of child node 
	 {\tt N}; similarly, {\bf llvm.hpvm.bind.output}}\\
\hline
\end{tabular}
%------------------------------------------------------------------------------
\begin{tabular}{p{0.5\textwidth}|p{0.5\textwidth}}
\multicolumn{2}{l}{\em Intrinsics for Querying Graphs}\\
\hline
{\tt\footnotesize i8* {\bf llvm.hpvm.getNode}()}	&
	{\small Return a handle to the current dataflow node}\\
{\tt\footnotesize i8* {\bf llvm.hpvm.getParentNode}(i8* N)}:	&
	{\small Return a handle to the parent of node \texttt{N}}\\
{\tt\footnotesize i32 {\bf llvm.hpvm.getNodeInstanceID.[xyz]}(
                       \hspace*{2em}i8* N)}	&
	{\small Get index of current dynamic node instance of node N in 
	 dimension {\tt x}, {\tt y} or {\tt z}}.\\
{\tt\footnotesize i32 {\bf llvm.hpvm.getNumNodeInstances.[xyz]}(
                       \hspace*{2em}i8* N)}	&
	{\small Get number of dynamic instances of node \texttt{N}
    in dimension \texttt{x}, \texttt{y} or \texttt{z}}\\
{\tt\footnotesize i32 {\bf llvm.hpvm.getVectorLength}(i32 typeSz)}	&
	{\small Get vector length in target compute unit for type size \texttt{typeSz}}\\
\hline
\end{tabular}
%------------------------------------------------------------------------------
\begin{tabular}{p{0.5\textwidth}|p{0.5\textwidth}}
\multicolumn{2}{l}{\em Intrinsics for Memory Allocation and Synchronization}\\
\hline
{\tt\footnotesize i8* {\bf llvm.hpvm.malloc}(i32 nBytes)}	&
{\small Allocate a block of memory of size {\tt nBytes} and return pointer to it}\\
{\tt\footnotesize i32 {\bf llvm.hpvm.atomic.add}(i32*, i32),
~~~~~ \hspace*{3em}i32 {\bf llvm.hpvm.xchg}(i32, i32), $\ldots$} &
{\small Atomic-fetch-and-add, atomic-swap, etc., on shared memory locations}\\
{\tt\footnotesize void {\bf llvm.hpvm.barrier()}}:	&
{\small \emph{Local} synchronization barrier across dynamic instances of
current leaf node}\\
\hline
\end{tabular}
%------------------------------------------------------------------------------
\begin{tabular}{p{0.5\textwidth}|p{0.5\textwidth}}
\multicolumn{2}{l}{\em Intrinsics for Integration with Host Code}\\
\hline
{\tt\footnotesize i8* {\bf llvm.hpvm.launch}(Function* F, i8* args, i1 Stream)}	&
	{\small Launch graph associated with Function {\tt F} asyncronously}\\
{\tt\footnotesize i8* {\bf llvm.hpvm.wait}(i8* graphID)}	&
	{\small Wait for completion of graph {\tt graphID}}\\
{\tt\footnotesize i8* {\bf llvm.hpvm.push}(i8* graphID, i8* args)}	&
	{\small Push args as streaming input of graph {\tt graphID}}\\
{\tt\footnotesize i8* {\bf llvm.hpvm.pop}(i8* graphID)}	&
	{\small Read next streaming output of graph {\tt graphID}}\\
\hline
\end{tabular}
%------------------------------------------------------------------------------
}
\caption{Intrinsic functions used to implement the \NAME{} internal 
	representation.  {\tt i$N$} is the $N$-bit integer type in LLVM.}
\label{table:intrinsics}
\end{table*}
%------------------------------------------------------------------------------

\NAME{} is implemented on top of LLVM~\cite{LLVM:CGO04} using LLVM intrinsic
functions (this choice is explained further at the end of this section).
For a brief summary, refer to Table~\ref{table:intrinsics}.
We use \NAME{} intrinsics to describe and query
the structure of the dataflow graph - \emph{Graph Intrinsics} and
\emph{Query Intrinsics} respectively, with additional intrinsics for memory
allocation and synchronization.

Each dataflow node is associated with a function, specified as function pointer,
describing the functionality of the corresponding node.
The incoming dataflow edges correspond to the incoming arguments to the 
node function.
The outgoing dataflow edges are represented by 
the return type of the node function, which must be an LLVM struct type with
one field per outgoing edge.

Functions
associated with internal nodes may only contain \NAME{} graph intrinsics.
Listing~\ref{lst:hpvm} contains the internal node function {\tt SgemmInternal}
(lines 46-65), associated with dataflow node {\tt SgemmInternal}.
This function only includes \NAME{} graph intrinsics, e.g. to create the leaf
nodes {\tt Allocation} (line 50) and {\tt SgemmLeaf} (line 51) and the dataflow
edges between them.

Functions
associated with leaf nodes contain llvm IR with vector instructions and may also
contain \NAME{} query, allocation and synchronization intrinsics.
Listing~\ref{lst:hpvm} contains the leaf node functions {\tt Allocation}
(lines 14-23) and {\tt SgemmLeaf} (lines 26-43), associated with dataflow nodes
{\tt Allocation} and {\tt SgemmLeaf}. Note the use of \NAME{} query and
allocation intrinsics, in lines 19, 31-37. Also, that the return type of
function {\tt Allocation} (defined in line 7) has two elements, just as the
number of outgoing dataflow edges, while function {\tt SgemmLeaf}, with no
outgoing edges, has an empty struct (defined in line 9) as a return type.

Finally, we define intrinsics aimed to
integrate the dataflow graph in the host code (which will be performing the
computation that is not to be mapped to accelerators, e.g. initialization,
output).
In Listing~\ref{lst:hpvm}, the host code initiates the execution of a dataflow
graph associated with the function {\tt SgemmRoot} and waits for its completion
(lines 102 and 103 respectively).

The following listing includes a description of the \NAME{}
intrinsics, divided according to their functionality. LLVM type i8* is used as 
an opaque handle to represent dataflow nodes and edges in the instruction set.
\begin{enumerate}
\item \NAME{} Graph Intrinsics: describing the structure of the dataflow graph:
\begin{itemize}
\item {\tt i8* llvm.hpvm.createNode(Function * F)}: Creates a dataflow node
      associated with the function {\tt F}
\item {\tt i8* llvm.hpvm.createNode\{1-3\}D(Function * F, int $n_1$, ...)}:
      Creates $n_1$ dataflow nodes, all associated with the function {\tt F}.
      There are three versions of this intrinsic depending on the number of
      dimensions of the grid of nodes.
\item {\tt void llvm.hpvm.createEdge(i8* Src, i8* Dst, i1 ReplTy, i32 sp, i32 dp, i1 Stream)}:
      Creates a dataflow edge from output $sp$ of node {\tt Src} to ipput $dp$
      of node {\tt Dst}. {\tt Stream} argument specifies whether the defined
      dataflow edge is an ordinary or a streaming edge. {\tt ReplTy} specifies
      the replication type, ''one-to-one'' or ''all-to-all''.
\item {\tt void llvm.hpvm.bind.input(i8* N, i32 ip, i32 ic, i1 Stream)}: Bind
      input $ip$ of node {\tt N}'s parent node to input $ic$ of child node
      {\tt N}.
\item {\tt void llvm.hpvm.bind.output(i8* N, i32 oc, i32 op, i1 Stream)}: Bind
      output $oc$ of node {\tt N} to output $op$ of its parent node.
\end{itemize}
\item \NAME{} Query Intrinsics: querying the structure of the dataflow
       graph:
\begin{itemize}
\item {\tt i8* llvm.hpvm.getNode()}: Get a pointer to the current dataflow node
\item {\tt i8* llvm.hpvm.getParentNode(i8* N)}: Get a pointer to the parent
      node, in the dataflow graph hierarchy, of node {\tt N}
\item {\tt i32 llvm.hpvm.getNumDims(i8* N)}: Get the number of dimensions of the
      grid of node {\tt N}
\item {\tt i32 llvm.hpvm.getNodeInstanceID.[xyz](i8* N)}: Get a unique
      identifier of a node instance of node {\tt N} in the specified dimension
\item {\tt i32 llvm.hpvm.getNumNodeInstances.[xyz](i8*, i32)}: Get the number
      of dynamic instances of node {\tt N} in the specified dimension
\item {\tt i32 llvm.hpvm.getVectorLength(i32 typeSize)}: Get vector length in
      target compute unit for type size {\tt typeSize}
\end{itemize}
\item \NAME{} Memory Allocation Intrinsics:
\begin{itemize}
\item {\tt i8* llvm.hpvm.malloc(i32 nBytes)}: Allocates an object of size
      $nBytes$ in global memory, shared by all nodes, although the pointer
      returned must somehow be communicated explicitly (just as with heap
      allocation in a multi-threaded C program).
\end{itemize}
\item \NAME{} Synchronization Intrinsics:
\begin{itemize}
\item {\tt void llvm.hpvm.barrier()}: Only synchronizes the dynamic instances
      of the node that executes it, and not all other concurrent nodes.
\item Atomic operations, e.g. {\tt i32 llvm.hpvm.atomic.add(i32*, i32)}.
\end{itemize}
\item \NAME{} Integration with Host Intrinsics:
\begin{itemize}
\item {\tt i8* llvm.hpvm.launch(Function* F, i8* args, i1 Stream)}: Launch
      dataflow graph associated with Function {\tt F} asyncronously, and return
      a handle to identify the graph. Argument {\tt Stream} indicates whether
      the dataflow graph contains streaming dataflow edges.
\item {\tt void llvm.hpvm.wait(i8* graphID)}: Wait for completion of graph
      {\tt graphID}.
\item {\tt void llvm.hpvm.push(i8* graphID, i8* args)}: Push args as streaming
      input of graph {\tt graphID}. {\tt args} is a packed struct containing all
      arguments expected by the executing graph.
\item {\tt i8* llvm.hpvm.pop(i8* graphID)}: Read streaming output of graph
      {\tt graphID}.
\end{itemize}
\end{enumerate}

-
\NAME{} is implemented on top of LLVM~\cite{LLVM:CGO04} using LLVM intrinsic
functions. This approach has many advantages. It is not-intrusive, allowing
smooth integration existing analysis optimization passes, while at the same
time enabling our custom passes to recognize them and handle each according
to their specified semantics. The memory behaviour of the intrinsic functions
can be specified, thus providing existing passes with the information required
to include them in their analysis or transformation.

Additionally, using intrinsics we easily isolate the parallel code from the code
executing sequentially (LLVM IR) ; we use the intrinsic\\
{\tt llvm.hpvm.createNode**(Function* F, ...)} and the parallel code is
included in {\tt F}. This has tha additional benefits of easily identifying the
communication between the sequentially and parallely executing parts. Only what
is passed as arguments to the dataflow graph is accesible by it. This simplifies
code generation in that generating code for the node function {\tt F} alone is
sufficient.

%------------------------------------------------------------------------------
\subsection{Performance features}
\label{sec:design:performance}
%------------------------------------------------------------------------------
When global memory must be shared across nodes mapped to devices
with separate address spaces, the translator inserts calls to the
appropriate accelerator runtime API (e.g., the OpenCL run-time) to perform 
the copies.
Such copies are sometimes redundant, e.g., if the data has already been
copied to the device by a previous node execution. It is important to avoid
unnecessary memory copies between devices for good performance.
We achieve this in two ways.

First, we differentiate between pointers to input/output data arrays using 
attributes {\tt in}, {\tt out}, and {\tt inout} to node arguments.
For example, annotating a pointer argument of a node as {\tt out} allows us to
avoid copying its initial data to the compute unit for which that node is
compiled.
In Listing~\ref{lst:hpvm}, matrices $A$ and $B$ are annotated as {\tt in}
arguments, and thus are only inputs to the dataflow node {\tt SgemmLeaf}, but
$C$ is both input and output (line 26).

Second, we implement a conceptually simple ``memory tracker''
to record the locations of the latest copy of data arrays, and uses this to 
avoid unnecessary copies.
Calls to the memory tracker due to dataflow edge semantics are automatically
inserted to the generated code. Host code, however, is required to explicitly
use the memory tracker interface, since it does not access data using dataflow
edges.
The memory tracker runtime calls are:
\begin{itemize}
\item {\tt llvm\_hpvm\_track\_mem(i8* ptr, i64 memsize)}: Add the memory region
      starting from pointer {\tt ptr} of size {\tt memsize} in the tracked
      memory table.
\item {\tt llvm\_hpvm\_request\_mem(i8* ptr, i64 size)}: Request to access the
      memory region associated woth pointer {\tt ptr}. This automatically
      results in a data copy if data is not already present in the host memory.
\item {\tt llvm\_hpvm\_untrack\_mem(i8* ptr)}: Remove the memory region
      assiosiated with pointer {\tt ptr} in the tracked memory table.
\end{itemize}

In Listing~\ref{lst:hpvm}, the host code contains calls to the \NAME{} runtime
denoting that the locations of matrices $A$, $B$ and $C$ should be tracked (or
no longer tracked), in lines 89-91 (109-111 respectively). Also, the host
explicitly asks for the matrix $C$ to be available in host memory (line 106).

%------------------------------------------------------------------------------
\section{Compilation Strategy}
\label{sec:compilation}
%------------------------------------------------------------------------------

%------------------------------------------------------------------------------
\subsection{Frontend}
\label{sec:compilation:frontend}
%------------------------------------------------------------------------------
The front end parses the source files and generates the textual representation
of \NAME{}, a hierarchical dataflow graph (DFG) represented through \NAME{}
intrinsics in LLVM IR along with code that may include \NAME{} intrinsics for
the leaf nodes.
Currently, we use C with dummy function calls with one-to-one correspondence to
\NAME{} intrinsics to generate \NAME{}.

%------------------------------------------------------------------------------
\subsection{Analysis}
\label{sec:compilation:analysis}
%------------------------------------------------------------------------------
First, the Graph Builder Pass constructs an internal representation for the DFG
by parsing the \NAME{} intrinsics.
Then, other passes may operate on and optimize the LLVM IR in the leaf nodes.
Also, we have built analysis passes that determine features of interest in the
DFG, e.g. read only memory (constantMemPass) or memory that is private and
visible only to a set of dataflow nodes (localMemPass). This inormation is used
for efficient code generation.

%------------------------------------------------------------------------------
\subsection{Backend Code Generation}
\label{sec:compilation:backend}
%------------------------------------------------------------------------------
In general, code generation proceeds in a bottom up approach, maintaining the
invariant that when a node is encountered as a child node, code generation for
it has already been performed and a native function that performs the node's
computation, {\tt genFunc}, has been generated. {\tt genFunc} is exposed to the
higher graph levels as an interface to invoke the node's execution.

The code generation pass uses the static dataflow graph to:

\begin{itemize}
\item Identify nodes which can be mapped to one or more available compute
      units efficiently. Currently, we use user-provided hints as to which
      target device to compile for, and thus which backend to invoke, e.g. in
      Listing~\ref{lst:hpvm} {\tt SgemmLeaf} node is annotated as a GPU node
      using metadata (lines 119, 122).
\item For the identified node, invoke appropriate backend to generate kernel
      code for the chosen target, and calls to \NAME{} runtime.
      The \NAME{} runtime invokes
      the appropriate accelerator runtime API to launch the generated kernel.
      All these calls are encapsulated in {\tt genFunc}.
\end{itemize}

At the end of code generation, we have host code with \NAME{} runtime calls and
target specific kernels.

Currently, we have implemented backend translators for three targets: nVidia
GPUs, Intel Vector Hardware, and x86 processors. We leverage existing
infrastructure for some of the translation process, the LLVM x86 and NVPTX
backends and the proprietary Intel OpenCL compiler.

Listings~\ref{lst:kernel} and~\ref{lst:host} show fragments of the kernel and
host code generated by the GPU backend respectively. For kernel code generation,
in Listing~\ref{lst:kernel}, note that the \NAME{} intrinsics (previously in
lines 31-37 of Listing~\ref{lst:hpvm}) have been replaced by function calls to
OpenCL library functions (please refer to associated code and comment in
Listing~\ref{lst:kernel}, lines 13-20).

For host code generation, in Listing~\ref{lst:host}, note the function
{\tt SgemmRoot2}, defined in lines 76-143.
This is the function generated by the GPU backend that encaptulates the
generated calls to OpenCL runtime that are required to setup and call the
generated kernel.

%------------------------------------------------------------------------------
\subsection{Code Generation for Memory Tiling}
\label{sec:compiler:tiling}
%------------------------------------------------------------------------------

If the programmer wished to ``tile'' the computation, she would use an
additional level in the dataflow graph hierarchy.
The (1D, 2D or 3D) instances of a leaf node would become a single (1D, 2D or
3D) tile of the computation.
The (1D, 2D or 3D) instances of the parent node of the leaf node would become 
the (1D, 2D or 3D) blocks of tiles.
Thus, \emph{a single mechanism, the hierarchical dataflow graph, represents both
tiling for scratchpad memory on the GPU and tiling for cache on the CPU}.
On a GPU, the leaf node becomes a thread block and we create
as many thread blocks as the dimensions of the parent node.
On a CPU or AVX target, the code results in a loop nest with
as many blocks as the dimensions of the parent node, of tiles as large
as the dimensions of the leaf node.

Any memory allocated for each tile (using {\tt llvm.hpvm.malloc} intrinsic in
a leaf node) would be assigned to scratchpad memory on
a GPU or left in global
memory and get transparently cached due to temporal reuse on the CPU.
We refer to a node performing such
allocations as {\em allocation node} due to its functionality.

We have used this mechanism to create tiled versions of
four of the seven
Parboil benchmarks evaluated in Section~\ref{sec:evaluation}.
The tile sizes are determined by the programmer in our experiments. The
expectation is that they will usually be determined by upstream optimization
passes or autotuning tools.
For the three
benchmarks ({\tt sgemm, tpacf, bfs}) for which non-tiled versions were available,
the tiled versions achieved a mean speedup of 19x on GPU and 10x on AVX,
with {\tt sgemm} getting as high as 31x speedup on AVX.

Figure~\ref{fig:sgemmdfg} shows an allocation node, {\tt Allocation}.
Listing~\ref{lst:hpvm} includes the node function (defined in lines 14-23),
that uses the {\tt llvm.hpvm.malloc} intrinsic to allocate memory and then
passes it to {\tt SgemmLeaf}. This memory, the argument {\tt \%shB}, is now
accesible only by the dynamic instances of {\tt SgemmLeaf} that are created by
the same dynamic instance of {\tt SgemmInternal}.
Listing~\ref{lst:kernel} shows that {\tt \%shB} has been mapped to the
scratchpad by the backend code generator, as shown by the address space
qualifier 3 (line 8).

%------------------------------------------------------------------------------
\subsection{Handling Pipelining}
\label{sec:compilation:pipeline}
%------------------------------------------------------------------------------
If a dataflow graph is identified as ''streaming'', then the dataflow nodes
need to be persistent: instead of performing a one-time computation and
complete their execution, they need to be available to process any new data
items transfarred through the streaming edges until the data stream ends.

For each pipeline stage, we create a seperate thread to perform its computation,
again represented by the native function {\tt genFunc} that was generated for
the assosiated node during code generation. The thread handles the required
synchronization between its stage and the other pipeline stages-threads, as well
as the data transfers represented by dataflow edges. Streaming edges are
implemented using buffering.

%------------------------------------------------------------------------------
\section{Compiler Optimization}
\label{sec:opt}
%------------------------------------------------------------------------------

An important capability of a compiler IR is to 
support effective compiler optimizations.
The hierarchical dataflow graph abstraction enables optimizations of explicitly
parallel programs at a higher (more informative) level of abstraction than
a traditional IR, like LLVM and many others, 
that lacks explicitly parallel abstractions.
We describe some of the primitive graph transformations enabled by the
representation, and two potentially important optimizations based on them --
fusion of compute kernels and mapping data to constant memory in GPUs.
Our long term goal is to develop a full-fledged parallel compiler 
infrastructure that leverages the parallel abstractions in \NAME{}.

%------------------------------------------------------------------------------
\subsection{Primitive Graph Operations}
\label{sec:opt:primitive}
%------------------------------------------------------------------------------

We use the graph abstractions in \NAME{} to implement a number of primitive
parallel program transformations as graph operations.
We define an {\it AllocCompute} graph as a simple two-node graph with one 
allocation node and one compute node.
This structure is needed when the dynamic instances of the compute node must 
share a single memory object, e.g., by reading and writing to parts of it.
\begin{enumerate}
\item
{\it MergeIndependentNodes}(N1, N2): Merge two parallel leaf nodes $N1$ and $N2$
with no path of edges connecting them into a single leaf node $N$. Make the
incoming edges of both $N1$ and $N2$ the incoming edges of $N$
(analogous, the outgoing edges).
$N1$ and $N2$ must have the same parent node,
dimensions, and size in each dimension.
\item
{\it MergeDependentNodes}(N1, N2): Like {\it MergeIndependentNodes}, 
except that $N1$ and
$N2$ are connected by one or more dataflow edges, which must all be 1-1 edges.
These edges are omitted from $N$,
and simply replaced with variable assignments (copies).
\item
{\it MergeAllocComputeGraphs}(N1, N2): Merge two internal nodes N1 and N2, each 
of which contains an AllocCompute graph, defined above, into a new internal node
$N$ containing a merged AllocCompute graph.  The allocation nodes of $N1$ and
$N2$ are merged into a new allocation node in $N$ and likewise for the compute
nodes.  There must be no path of edges from the compute node of $N2$ to the 
allocation node of $N1$ - that would prevent the merge of the allocation
nodes.
\item
{\it InlineAuxFunction}(F\_node, F\_aux): Inline an auxiliary function F\_aux
called from a node function F\_node.  This allows merging of $N1$
and $N2$ into a simple, new node function that calls the two original node
functions, without introducing additional overhead for the function calls and
without blocking optimizations across the function call boundaries.
%In our prototype, it also keeps the implementation of some intrinsics, like
%{\tt getParent} and {\tt getNodeInstance*} simple.
%
\end{enumerate}
In addition, note that the basic intrinsics, {\tt createNode*},
{\tt createEdge*}, {\tt bind.input}, {\tt bind.output}, {\tt 
getNodeInstanceID.*}, etc., are directly useful for many graph
analyses and transformations.

%------------------------------------------------------------------------------
\subsection{Node Fusion Using the Primitives}
\label{sec:opt:fusion}
%------------------------------------------------------------------------------

One key optimization we have implemented using these primitives is
{\em Node Fusion},
which can lead to more effective redundancy elimination across
kernels, reduced launch overhead when kernels are executed
on GPUs, improved temporal locality when kernels that reuse data are fused,
and sometimes reduced barrier synchronization overhead as well.
Merging nodes willy-nilly, however, can hurt performance greatly on some 
devices because of resource constraints or functional limitations.
For example, each thread-block in a GPU has limited memory and scratchpad 
capacity and merging two nodes into one could 
force the use of fewer thread blocks, reducing parallelism.
We use a simple policy to decide when to merge two nodes; for our experiments,
we augment this with manual annotations to 
specify which nodes should be merged.
We leave it to future work to develop a more sophisticated merging policy, 
perhaps guided by profile information or autotuning.

The algorithm we use attempts to merge two nodes if they have the same 
parent node, same target, and are both annotated for merging.
If so, the algorithm simply invokes the appropriate merge operation in the 
previous section (which check additional conditions the nodes must satisfy).

\todo{restore alrogithm for node fusion?}
% ****** FIXME: DEFINITELY RESTORE IF SPACE *****
%Algorithm~\ref{algo:loopfusion} shows the algorithm we use.
%
%%----------------------------------------------------------------
%\begin{algorithm}
%  \caption{Loop Fusion Algorithm}
%  \label{algo:loopfusion}
%  \begin{algorithmic}[1.]
%    \Procedure{LoopFusion}{Node $N1$, $N2$}: Node
%    \If{($N1$ and $N2$ have same parent node\\
%\hspace*{4em}         {\bf and} $N1$ and $N2$ have same target)}
%        \If{($N1$ and $N2$ are independent)}
%  	       \State $N$ = {\it MergeIndependentNodes}(N1, N2)
%        \Else
%          \If{(a single edge $E$ connects $N1$ to $N2$\\
%\hspace*{6em}				{\bf and} $E$ is a 1-1 edge)}
%  	       \State $N$ = {\it MergeDependentNodes}(N1, N2)
%	      \EndIf
%	    \EndIf
%    \EndIf
%    \State \Return N 
%    \EndProcedure
%  \end{algorithmic}
%\end{algorithm}
%%----------------------------------------------------------------

%------------------------------------------------------------------------------
\subsection{Mapping Data to GPU Constant Memory}
\label{sec:opt:constantmemory}
%------------------------------------------------------------------------------

GPU global memory is highly optimized (in nVidia GPUs) for coalescing of
consecutive accesses by threads in a thread block: irregular accesses can
have orders-of-magnitude lower performance.
In contrast, constant memory is optimized for read-only data 
that is invariant across threads and is much more efficient for
thread-independent data.

The \NAME{} translator for GPUs automatically identifies data that should be
mapped to constant memory.
The analysis is trivial for scalars, but also simple for array accesses
because of 
the \NAME{} intrinsics: for array index calculations, we identify  whether they 
depend on (1) the {\tt getNodeInstanceId.*}
intrinsics, which is the sole mechanism to express thread-dependent accesses,
or (2) memory accesses.
Those without such dependencies are uniform and are mapped to constant memory,
and the rest to GPU global memory.
\NAME{} translator idenified such candidates in 3 (spmv, tpacf, cutcp) out of 7 benchmarks
, resulting in 34\% performance improvement in tpacf and no
effect on performance of the other two benchmarks.

%------------------------------------------------------------------------------
\section{Evaluation}
\label{sec:evaluation}
%------------------------------------------------------------------------------

We evaluate the \NAME{} virtual ISA and compiler IR
by examining several questions:
(1) Is \NAME{} performance-portable: can we use the \emph{same virtual object 
code} to get ``good'' speedups on different compute units?
(2) Can \NAME{} achieve performance
competitive with hand-written OpenCL programs on our target architectures?
(3) Does \NAME{} effectively capture pipelined parallelism for streaming
computations, which cannot be easily expressed in existing alternatives
like PTX, HSAIL and SPIR?

%------------------------------------------------------------------------------
\subsection{Experimental Setup and Benchmarks}
\label{sec:evaluation:setup}
%------------------------------------------------------------------------------
Starting from C with a set of functions corresponding to the \NAME{} intrinsics
we write parallel applications equivalent to a set
of OpenCL applications, as described in section~\ref{sec:compilation}.
We translated the \emph{same} \NAME{} code to two different target units:
the AVX instruction set in an Intel Xeon E5 core i7 and 
a discrete nVidia GeForce GTX 680 GPU card with 2GB of memory.
The Intel Xeon also served as the host processor, running
at 3.6 GHz, with 16 GB RAM.

We used seven applications from the 
Parboil benchmark suite~\cite{Parboil}:
Sparse Matrix Vector Multiplication (spmv), 
Single-precision Matrix Multiplication (sgemm), 
Stencil PDE solver (stencil), 
Lattice-Boltzmann (lbm), 
Breadth-first search (bfs), 
Two Point Angular Correlation Function (tpacf), and
Distance-cutoff Coulombic Potential (cutcp).
For the last question, we used a pipelined streaming application,
described later below.

In the GPU experiments, our baseline for comparison is the best available 
OpenCL implementation. For spvm, sgemm, lbm, bfs and cutcp, that is the
Parboil version labeled {\tt opencl\_nvidia}, which 
has been hand-tuned for the Tesla NVidia GPUs~\cite{Liwen:Personal}.
For stencil and tpacf, the best is the generic Parboil version labeled
{\tt opencl\_base}. We further optimized the codes by removing unnecessary data
copies (bfs) and global barriers (tpacf, cutcp).
nVidia's proprietary OpenCL compiler is used to compile all applications.

In the vector experiments, with the exception of bfs, our baseline is the same
OpenCL implementations we chose as GPU baselines, 
but compiled using the Intel OpenCL compiler, because these achieved
the best vector performance as well.
For bfs, we used
{\tt opencl\_base} instead, as {\tt opencl\_nvidia} failed the correctness test.
The \NAME{} versions were generated to match the algorithms used in the OpenCL
versions, \emph{and that was used for both vector and GPU experiments}.

We use the largest available input for each benchmark, and each data point we
report is an average of ten runs. 
or generate a larger
input when the runtime would be too small.
Each data point we report is an average of ten runs.
We repeated the experiments multiple times to verify their stability

\begin{figure*}[hbt]
\begin{minipage}{0.48\textwidth}
\begin{center}
    \includegraphics[height=4cm]{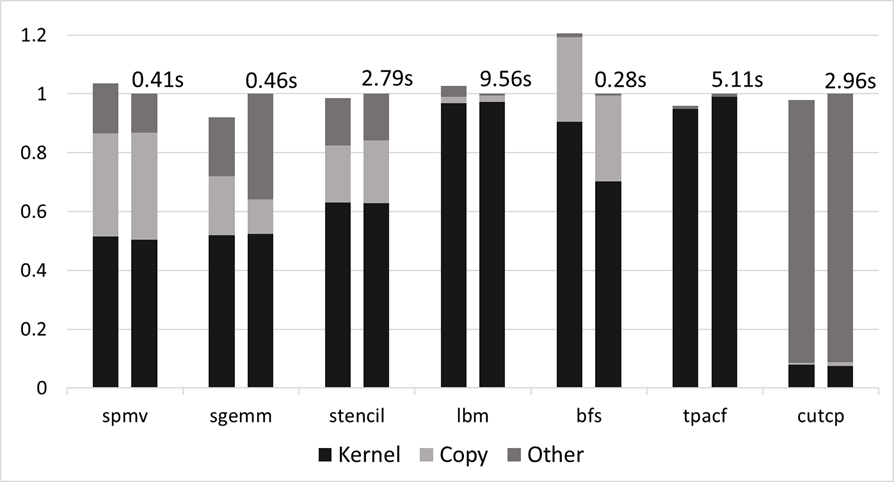}
    \caption{\footnotesize{GPU Experiments - Normalized Execution Time. For each
    benchmark, left bar is \NAME{} and right bar is OpenCL.}}
%    \caption{\footnotesize{GPU Experiments - Large Test Normalized Execution
%    Time}}
    \label{fig:gpularge}
\end{center}
\end{minipage}~~~~\begin{minipage}{0.48\textwidth}
\begin{center}
    \centering
    %\hspace*{4ex}
    \includegraphics[height=4cm]{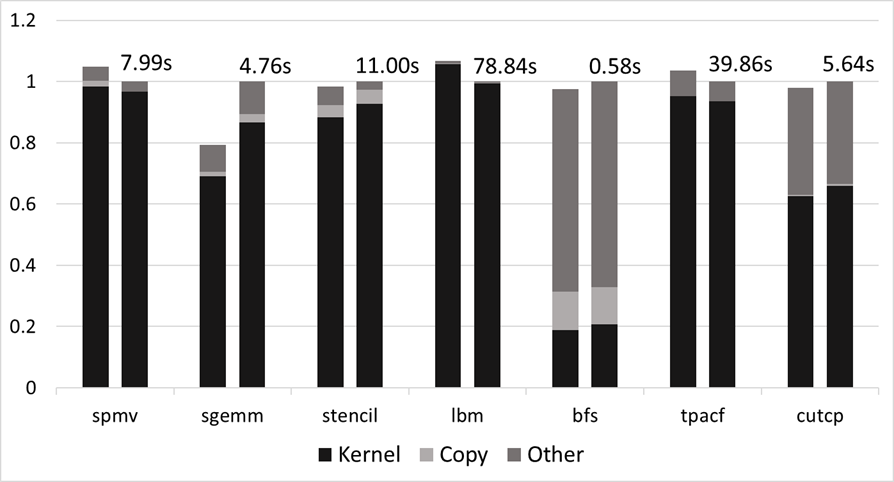}
%    \caption{\footnotesize{Vector Experiments - Large Test Normalized Execution
%    Time}}
    \caption{\footnotesize{Vector Experiments - Normalized Execution Time. For
    each benchmark, left bar is \NAME{} and right bar is OpenCL.}}
    \label{fig:cpularge}
\end{center}
\end{minipage}
\end{figure*}

Figures~\ref{fig:gpularge} and~\ref{fig:cpularge}
show the execution
time of these applications on GPU and vector hardware respectively,
normalized to the baselines mentioned above.
Each bar shows segments for the 
time spent in the compute kernel (kernel), 
copying data (copy) 
and remaining time on the host.
The total execution time for the baseline is shown above the 
bar to indicate the actual running times.

Comparing with the GPU baseline, \NAME{} achieves near
hand-tuned OpenCL performance for all benchmark except bfs,
where \NAME{} takes 20\% longer.  The overhead is
because our translator is not mature enough to generate
global barriers, and thus \NAME{} issues more kernels than
the {\tt opencl\_nvidia} version, incurring significant overhead.

In the vector case, \NAME{} is within 7\% of the hand-tuned baseline in the
worst case.
We see 20\% speedup for
sgemm, resulting exclusively from the kernel execution time, 
because LLVM applies more effective optimizations on the kernel than
the nVidia OpenCL compiler.

\todo{add paragraph about less information compared to OpenCL?}
\comment{I did not include paragraph about vector length in benchmarks}

%------------------------------------------------------------------------------
\subsection{Evaluation of Pipelined Parallelism}
\label{sec:evaluation:streaming}
%------------------------------------------------------------------------------

\begin{figure}[hbt]
\begin{center}
  \includegraphics[height=4cm]{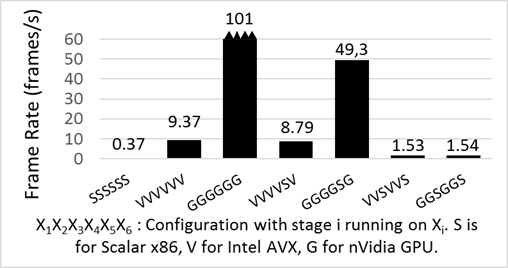}
  \caption{\footnotesize{Frame rates of different
         configurations of Edge Detection six stage pipeline through single
         \NAME{} object code.}}
  \label{fig:pipelineResult}
\end{center}
\end{figure}

We used a six-stage image processing pipeline, Edge Detection in grey scale images, to
evaluate the benefits of pipelined parallelism in \NAME{}.
The application accepts a stream of grey scale images, $I$, and a fixed
mask $B$ and computes a stream of
binary images, $E$, that represent the edges of $I$. We feed 1280x1280 pixel
frames from a video as the input and measure the frame rate at the output.
This pipeline is natural to express in \NAME{}. The streaming edges and pipeline
stages simply map to key features of \NAME{}.
In contrast, expressing pipelined streaming parallelism in OpenCL, PTX, SPIR or
HSAIL, although possible, is extremely awkward. In particular, while the
parallel stages can be expressed using different
command queues, the (buffered) data transfers and synchronization between
stages are tedious, error-prone and difficult to scale to larger and more
complex pipelines.

Expressing this example in \NAME{} allows for flexibly
mapping computationally heavy stages of the pipeline to accelerators.
We can map each stage to
one of three targets (GPU, vector or a CPU thread), for a total of
$3^{6}=729$ different configurations, all generated from a single \NAME{} code.
Figure~\ref{fig:pipelineResult} shows the frame rate of 7 such
configurations.  
It shows that \NAME{} (a) can capture pipelined, streaming
computations effectively and achieve considerable speedups,
and (b) is flexible enough to allow a wide range of
configurations with different performance characteristics from
a single code.
The graph illustrates the dramatic
differences in performance between different mappings.
This flexibility of \NAME{} enables a (future) run-time scheduler to choose
configurations based on properties of the code, available hardware resources,
and energy constraints.

Note that, although our model allows us to exploit the pipeline parallelism
between different stages, we do not use the stream interface in our prototype
implementation. Thus the execution of different stages is not actually
overlapped when executing on a single device. We do get the benefit of
pipeline parallelism when we dispatch the execution of different stages to
multiple devices.

%------------------------------------------------------------------------------
\subsection{Node Fusion Optimization Evaluation}
\label{sec:evaluation:streaming}
%------------------------------------------------------------------------------

\begin{figure}[hbt]
\begin{center}
  \includegraphics[height=4cm]{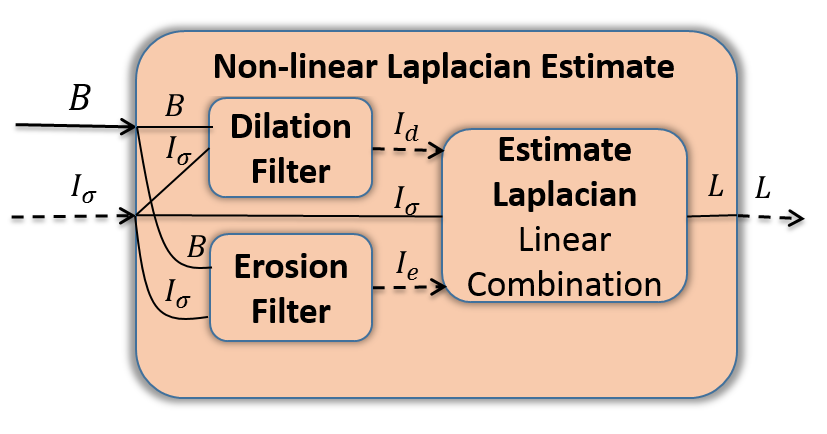}
  \caption{\footnotesize{Dataflow graph of Laplaciai Estimate.}}
  \label{fig:laplacian}
\end{center}
\end{figure}

We evaluated the benefits of Node Fusion using two widely used image processing
kernels, Laplacian Estimate ($L$) and Gradient Computation ($G$).
Most benchmarks we examined have been hand-tuned to
apply such transformations
manually, making it hard to find Node Fusion opportunities
(although they may often be more natural to write without manual merging).
The two kernels' dataflow graphs have similar structure, shown for $L$
in Figure~\ref{fig:laplacian}.
We compiled the codes to run entirely on GPU and
fed the same video frames as before.
We experimented with different
choices of node sets to merge.
Fusing all three nodes reduces the GPU invocations by two thirds,
but showed a slowdown for both $L$ and $G$.
Fusing just the two independent nodes
reduced the number of GPU invocations by one third and gave a speedup of
7.4\% and 10.4\% on $L$ and $G$ respectively. These experiments show
that significant benefits can occur 
when Node Fusion is applied to a carefully selected set of dataflow nodes.

%------------------------------------------------------------------------------
\section{Example}
\label{sec:example}
%------------------------------------------------------------------------------

In this section, we show an example in \NAME{} and the translation process for
an nVidia GPU target.
We use the sgemm benchmark from the Parboil test suite. 

%------------------------------------------------------------------------------
\subsection{SGEMM in HPVM}
\label{sec:example:hpvm}
\begin{figure}[hbt]
\begin{center}
  \includegraphics[height=7cm]{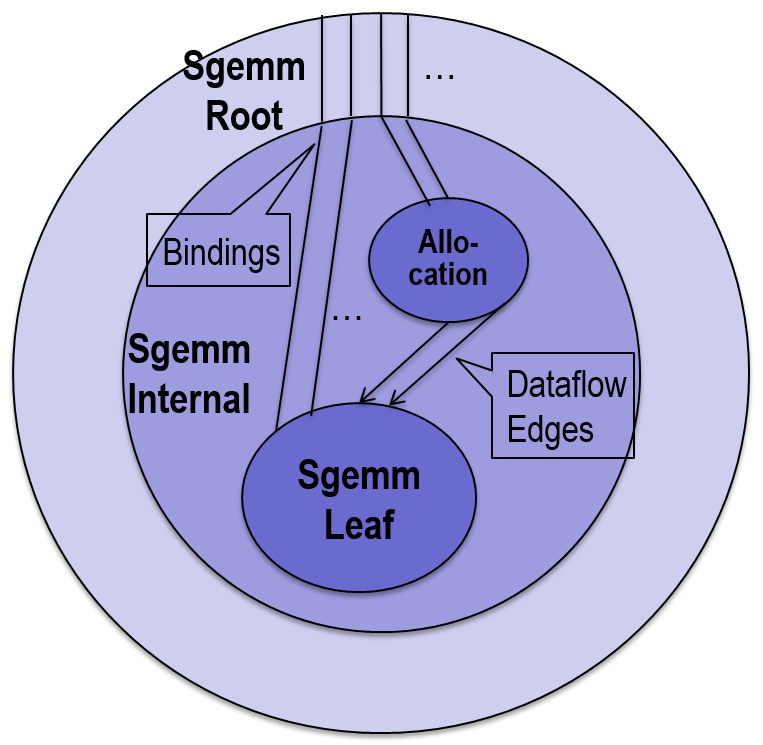}
  \caption{\footnotesize{Dataflow graph of sgemm.}}
  \label{fig:sgemmdfg}
\end{center}
\end{figure}

%------------------------------------------------------------------------------
In \NAME{}, this benchmark is
represented as a two level hierarchy with two leaf nodes performing memory
allocation and computation, Allocation and SgemmLeaf respectively, an internal
node SgemmInternal, plus a top level node, SgemmRoot, that interacts with the
host code. The additional level of hierarchy exists because the top level node
is required to have a single dynamic instance and is used to create the
underlying graph structure.
This is shown in figure~\ref{fig:sgemmdfg}. Listing~\ref{lst:hpvm} shows
parts of the \NAME{} code that represents this dataflow graph, with inlined
comments to point out the performed operations.

\begin{lstlisting}[language=llvm,
                   label=lst:hpvm,
                   caption=SGEMM in HPVM]
; ModuleID = 'build/hpvm_sh_default/main.hpvm.ll'

; %struct.RootIn is the struct type used to pack the arguments required by the dataflow graph
%struct.RootIn = type <{ float*, i64, i32, float*, i64, i32, float*, i64, i32, i32, float, float, i32, i32, i32, i32 }>
; Examples of dataflow node return types:
; - Allocation Node returns a pointer and an i64
%struct.out.Allocation = type <{ i8*, i64 }>
; - SgemmLeaf Node does not return anything
%emptyStruct = type <{}>

; ...

; Allocation Node function, performing one of size block_x x block_y
define %struct.out.Allocation @Allocation(i32 %block_x, i32 %block_y) #2 {
entry:
  %mul = mul nsw i32 %block_y, %block_x
  %conv = sext i32 %mul to i64
  %mul1 = shl nsw i64 %conv, 2
  %call1 = call i8* @llvm.hpvm.malloc(i64 %mul1)
  %returnStruct = insertvalue %struct.out.Allocation undef, i8* %call1, 0
  %returnStruct2 = insertvalue %struct.out.Allocation %returnStruct, i64 %mul1, 1
  ret %struct.out.Allocation %returnStruct2
}

; Computation Leaf Node function, using HPVM query intrinsics
define %emptyStruct @SgemmLeaf(float* in %A, i64 %bytesA, i32 %lda, float* in %B, i64 %bytesB, i32 %ldb, float* in out %C, i64 %bytesC, i32 %ldc, i32 %k, float %alpha, float %beta, float* nocapture %shB, i64 %bytesshB) #2 {
entry:
  %c = alloca [16 x float], align 16
  %0 = bitcast [16 x float]* %c to i8*

  %call7 = call i8* @llvm.hpvm.getNode()
  %call18 = call i8* @llvm.hpvm.getParentNode(i8* %call7)
  %call29 = call i32 @llvm.hpvm.getNodeInstanceID.x(i8* %call7)
  %call310 = call i32 @llvm.hpvm.getNodeInstanceID.y(i8* %call7)
  %call411 = call i32 @llvm.hpvm.getNodeInstanceID.x(i8* %call18)
  %call512 = call i32 @llvm.hpvm.getNodeInstanceID.y(i8* %call18)
  %call613 = call i32 @llvm.hpvm.getNumNodeInstances.x(i8* %call7)

  %mul = mul nsw i32 %call613, %call310
  %add = add nsw i32 %mul, %call29
; ... Remaining Computation
  ret %emptyStruct undef
}

; Internal Node function, using HPVM graph intrinsics
define %emptyStruct.19 @SgemmInternal(float* in %A, i64 %bytesA, i32 %lda, float* in %B, i64 %bytesB, i32 %ldb, float* in out %C, i64 %bytesC, i32 %ldc, i32 %k, float %alpha, float %beta, i32 %block_x, i32 %block_y) #2 {
entry:

; Create two child nodes, Allocation and SgemmLeaf
  %Allocation.node = call i8* @llvm.hpvm.createNode(i8* bitcast (%struct.out.Allocation (i32, i32)* @Allocation to i8*))
  %SgemmLeaf.node = call i8* @llvm.hpvm.createNode2D(i8* bitcast (%emptyStruct (float*, i64, i32, float*, i64, i32, float*, i64, i32, i32, float, float, float*, i64)* @SgemmLeaf to i8*), i32 %block_x, i32 %block_y)

; Create the dataflow edges between nodes Allocation and SgemmLeaf
  %output = call i8* @llvm.hpvm.createEdge(i8* %Allocation.node, i8* %SgemmLeaf.node, i1 true, i32 0, i32 12, i1 false)
  %output1 = call i8* @llvm.hpvm.createEdge(i8* %Allocation.node, i8* %SgemmLeaf.node, i1 true, i32 1, i32 13, i1 false)

; Bindings to transfer its inputs to child nodes Allocation and SgemmLeaf
  call void @llvm.hpvm.bind.input(i8* %SgemmLeaf.node, i32 0, i32 0, i1 false)
  call void @llvm.hpvm.bind.input(i8* %SgemmLeaf.node, i32 1, i32 1, i1 false)
; ...
  call void @llvm.hpvm.bind.input(i8* %Allocation.node, i32 12, i32 0, i1 false)
  call void @llvm.hpvm.bind.input(i8* %Allocation.node, i32 13, i32 1, i1 false)

  ret %emptyStruct.19 undef
}

; Root Node function, using HPVM graph intrinsics
define %emptyStruct.20 @SgemmRoot(float* in %A, i64 %bytesA, i32 %lda, float* in %B, i64 %bytesB, i32 %ldb, float* in out %C, i64 %bytesC, i32 %ldc, i32 %k, float %alpha, float %beta, i32 %block_x, i32 %block_y, i32 %grid_x, i32 %grid_y) #2 {
entry:
; Creates internal node SgemmInternal
  %SgemmInternal.node = call i8* @llvm.hpvm.createNode2D(i8* bitcast (%emptyStruct.19 (float*, i64, i32, float*, i64, i32, float*, i64, i32, i32, float, float, i32, i32)* @SgemmInternal to i8*), i32 %grid_x, i32 %grid_y)

; Bindings to transfer its inputs to child node SgemmInternal
  call void @llvm.hpvm.bind.input(i8* %SgemmInternal.node, i32 0, i32 0, i1 false)
  call void @llvm.hpvm.bind.input(i8* %SgemmInternal.node, i32 1, i32 1, i1 false)
; ...
  ret %emptyStruct.20 undef
}

; Function launching the dataflow graph defined by SgemmRoot and
; performing calls to HPVM runtime
define i32 @main(i32 %argc, i8** %argv) #2 {
entry:
; ...
; Placeholder for HPVM runtime to perform required initializations
  call void @llvm.hpvm.init()
; Host calls to HPVM runtime, noting that the locations of arrays A, B,
; and C should be tracked
  call void @llvm_hpvm_track_mem(i8* %24, i64 %mul19) #1
  call void @llvm_hpvm_track_mem(i8* %26, i64 %mul22) #1
  call void @llvm_hpvm_track_mem(i8* %28, i64 %mul25) #1
; ...
; Argument Packing
  %call23 = tail call noalias i8* @malloc(i64 88) #1
  %A1.i = bitcast i8* %call23 to float**
  store float* %A, float** %A1.i, align 1, !tbaa !3
  %bytesA2.i = getelementptr inbounds i8* %call23, i64 8
  %23 = bitcast i8* %bytesA2.i to i64*
  store i64 %bytesA, i64* %23, align 1, !tbaa !6
; ... remaining arguments packed
; Launch dataflow graph SgemmRoot with arguments packed in %call23
  %graphID = call i8* @llvm.hpvm.launch(i8* bitcast (%emptyStruct.20 (float*, i64, i32, float*, i64, i32, float*, i64, i32, i32, float, float, i32, i32, i32, i32)* @SgemmRoot to i8*), i8* %call23, i1 false)
  call void @llvm.hpvm.wait(i8* %graphID)
; ...
; Request array C
  call void @llvm_hpvm_request_mem(i8* %37, i64 %mul25) #1
; ...
; Clear memory tracker table
  call void @llvm_hpvm_untrack_mem(i8* %39) #1
  call void @llvm_hpvm_untrack_mem(i8* %41) #1
  call void @llvm_hpvm_untrack_mem(i8* %43) #1
; Placeholder for HPVM runtime to perform cleanup
  call void @llvm.hpvm.cleanup()
; ...
  ret i32 0
}

; Hints for compilation target
!hpvm_hint_gpu = !{!0}
!hpvm_hint_cpu = !{!1, !2}

!0 = metadata !{%emptyStruct (float*, i64, i32, float*, i64, i32, float*, i64, i32, i32, float, float, float*, i64)* @SgemmLeaf}
!1 = metadata !{%emptyStruct.19 (float*, i64, i32, float*, i64, i32, float*, i64, i32, i32, float, float, i32, i32)* @SgemmInternal}
!2 = metadata !{%emptyStruct.20 (float*, i64, i32, float*, i64, i32, float*, i64, i32, i32, float, float, i32, i32, i32, i32)* @SgemmRoot}
\end{lstlisting}

%------------------------------------------------------------------------------
\subsection{SGEMM Kernel generated from GPU backend}
\label{sec:example:nvvm}
%------------------------------------------------------------------------------
After code generation for the GPU target, a GPU kernel is generated which is shown in listing~\ref{lst:kernel}. Again, inlined comments point out the performed operations and how they are introduced by the backend.

\begin{lstlisting}[language=llvm,
                   label=lst:kernel,
                   caption=SGEMM - Kernel code generated from HPVM: LLVM
+ OpenCL library calls]
; ModuleID = 'build/hpvm_sh_default/main.hpvm.ll'
;...

; Generated Kernel. Note the address space qualifiers: addrspace(1)
; represents global address space, addrspace(3) represents local address
; space accesible only by work items in the same work group. A tile of
; matrix B has been placed in local memory.
define void @SgemmLeaf(float addrspace(1)* in %A, i64 %bytesA, i32 %lda, float addrspace(1)* in %B, i64 %bytesB, i32 %ldb, float addrspace(1)* in out %C, i64 %bytesC, i32 %ldc, i32 %k, float %alpha, float %beta, float addrspace(3)* nocapture %shB, i64 %bytesshB) #0 {
entry:
  %c = alloca [16 x float], align 16
  %0 = bitcast [16 x float]* %c to i8*
; ...
; Calls to OpenCL library calls are generated by the backend, to implement
; the HPVM intrinsics. The appropriate function call to implement an HPVM
; intrinsic depends on the dataflow graph structure.
  %1 = call i32 @get_local_id(i32 0)
  %2 = call i32 @get_local_id(i32 1)
  %3 = call i32 @get_group_id(i32 0)
  %4 = call i32 @get_group_id(i32 1)
  %5 = call i32 @get_local_size(i32 0)

  %mul = mul nsw i32 %5, %2
  %add = add nsw i32 %mul, %1
; ... remaining computation
  ret void
}

; The OpenCL function implementations for the target platform are found by
; linking with libclc
declare i32 @get_local_id(i32)
declare i32 @get_group_id(i32)
declare i32 @get_local_size(i32)
declare void @barrier(i32)

; Appropriate metadata is generated
!opencl.kernels = !{!3}
!nvvm.annotations = !{!4}

!3 = metadata !{void (float addrspace(1)*, i64, i32, float addrspace(1)*, i64, i32, float addrspace(1)*, i64, i32, i32, float, float, float addrspace(3)*, i64)* @SgemmLeaf}
!4 = metadata !{void (float addrspace(1)*, i64, i32, float addrspace(1)*, i64, i32, float addrspace(1)*, i64, i32, i32, float, float, float addrspace(3)*, i64)* @SgemmLeaf, metadata !"kernel", i32 1}
\end{lstlisting}

As pointed out in~\ref{lst:kernel} lines 28-29, the OpenCL library function
implementations are found in libclc and are available after linking.
For an nVidia GPU target, they are implemented using llvm intrinsics that are
recognized by the LLVM NVPTX backend (and translated down to reading special
hardware registers).
Listing~\ref{lst:libclcfunc} shows such a function that is included in the
kernel file we have genarated after linking.

\begin{lstlisting}[language=llvm,
                   label=lst:libclcfunc,
                   caption=OpenCL library function implementation from libclc]
; ModuleID = 'build/hpvm_sh_default/main.hpvm.ll'

; Function Attrs: alwaysinline nounwind readnone
define linkonce_odr i32 @get_local_id(i32 %dim) #2 {
entry:
  switch i32 %dim, label %return [
    i32 0, label %sw.bb
    i32 1, label %sw.bb1
    i32 2, label %sw.bb2
  ]
sw.bb:                                            ; preds = %entry
  %0 = tail call i32 @llvm.ptx.read.tid.x()
  br label %return
sw.bb1:                                           ; preds = %entry
  %1 = tail call i32 @llvm.ptx.read.tid.y()
  br label %return
sw.bb2:                                           ; preds = %entry
  %2 = tail call i32 @llvm.ptx.read.tid.z()
  br label %return
return:                                           ; preds = %sw.bb2, %sw.bb1, %sw.bb, %entry
  %retval.0 = phi i32 [ %2, %sw.bb2 ], [ %1, %sw.bb1 ], [ %0, %sw.bb ], [ 0, %entry ]
  ret i32 %retval.0
}

; Function Attrs: nounwind readnone
declare i32 @llvm.ptx.read.tid.x() #3
; Function Attrs: nounwind readnone
declare i32 @llvm.ptx.read.tid.y() #3
; Function Attrs: nounwind readnone
declare i32 @llvm.ptx.read.tid.z() #3

; ... Other OpenCL library functions ...
\end{lstlisting}

%------------------------------------------------------------------------------
\subsection{SGEMM Host generated from GPU backend}
\label{sec:example:host}
%------------------------------------------------------------------------------
The GPU backend also generates host code to perform the necessary data copies and launch the generated kernel. The generated host code is shown in listing~\ref{lst:host}. Again, inlined comments point out the performed operations.

\begin{lstlisting}[language=llvm,
                   label=lst:host,
                   caption=SGEMM Host generated from GPU backend]
; ModuleID = 'build/hpvm_sh_default/main.hpvm.ll'

; Function launching the dataflow graph defined by SgemmRoot and
; performing calls to HPVM runtime
define i32 @main(i32 %argc, i8** %argv) #2 {
entry:
; ...
; The following two calls replace the llvm.hpvm.init().
; HPVM runtime initializes the OpenCL context.
  %12 = call i8* @llvm_hpvm_ocl_initContext(i32 2)
; This call loads the kernel from disk, compiles and creates a kernel
; object associated with the dataflow graph. The information required
; to launch this kernel are returned. 
  %graphSgemmLeaf = call i8* @llvm_hpvm_ocl_launch(i8* %FilenamePtr, i8* %KernelNamePtr)
; Global variable storing the returned information
  store i8* %graphSgemmLeaf, i8** @graphSgemmLeaf.addr
; ...
; Host calls to HPVM runtime. Locations of arrays A, B, and C
; should be tracked
  call void @llvm_hpvm_track_mem(i8* %25, i64 %mul19) #1
  call void @llvm_hpvm_track_mem(i8* %27, i64 %mul22) #1
  call void @llvm_hpvm_track_mem(i8* %29, i64 %mul25) #1
; ...
; Argument Packing
  %call23 = tail call noalias i8* @malloc(i64 88) #1
  %A1.i = bitcast i8* %call23 to float**
  store float* %A, float** %A1.i, align 1, !tbaa !3
  %bytesA2.i = getelementptr inbounds i8* %call23, i64 8
  %23 = bitcast i8* %bytesA2.i to i64*
  store i64 %bytesA, i64* %23, align 1, !tbaa !6
; ... remaining argument packing
; Runtime call replacing the llvm.hpvm.lanch intrinsic.
; The called function, LaunchDataflowGraph, is responsible for calling
; the x86 function generated by the GPU backend that launches the
; generated kernel.
  %graphSgemmRoot = call i8* @llvm_hpvm_x86_launch(i8* (i8*)* @LaunchDataflowGraph, i8* %call23)
  call void @llvm_hpvm_x86_wait(i8* %graphSgemmRoot)
; ...
; Runtime call to request array C
  call void @llvm_hpvm_request_mem(i8* %38, i64 %mul25) #1
; Remove entries from memory tracker
  call void @llvm_hpvm_untrack_mem(i8* %40) #1
  call void @llvm_hpvm_untrack_mem(i8* %42) #1
  call void @llvm_hpvm_untrack_mem(i8* %44) #1
; Read information associated with kernel and clear the OpenCL runtime
; information and context
  %45 = load i8** @graphSgemmLeaf.addr
  call void @llvm_hpvm_ocl_clearContext(i8* %45)
  ret i32 0
}

; Declarations of HPVM runtime functions
declare void @llvm_hpvm_track_mem(i8*, i64) #0
declare i8* @llvm_hpvm_ocl_launch(i8*, i8*)
declare void @llvm_hpvm_ocl_wait(i8*)
; ...

; Function associated with allocation node, modified by the backend to
; only compute and return the allocation size.
; To perform local memory allocation for a GPU, the size needs to be known
; at launch time and passed as an argument. To that end, we use the
; modified allocation node function to perform the size computation.
; We insert a call to it in the host code before the kernel launch.
define %struct.out.Allocation @Allocation1(i32 %block_x, i32 %block_y) #2 {
entry:
  %mul = mul nsw i32 %block_y, %block_x
  %conv = sext i32 %mul to i64
  %mul1 = shl nsw i64 %conv, 2
  %returnStruct = insertvalue %struct.out.Allocation undef, i8* null, 0
  %returnStruct2 = insertvalue %struct.out.Allocation %returnStruct, i64 %mul1, 1
  ret %struct.out.Allocation %returnStruct2
}

; x86 function for kernel setup, data copying and call.
; This is genFunc (see section Compilation) for the generated kernel
define %emptyStruct.20 @SgemmRoot2(float* %A, i64 %bytesA, i32 %lda, float* %B, i64 %bytesB, i32 %ldb, float* %C, i64 %bytesC, i32 %ldc, i32 %k, float %alpha, float %beta, i32 %block_x, i32 %block_y, i32 %grid_x, i32 %grid_y) {
entry:
; Read information about the kernel that needs to be launched
; We use it to determine which kernel to set arguments for
  %graph.SgemmLeaf = load i8** @graphSgemmLeaf.addr

; Setting up kernel arguments
; llvm_hpvm_ocl_argument_ptr/scalar/shared are wrappers around the OpenCL
; runtime call setKernelArgument, that pass the arguments as required.
; llvm_hpvm_ocl_argument_ptr additionally makes calls to the memory
; tracker to determine if memory copy is required for the passed
; argument. If required performs the data transfer and updates the
; memory tracker information.
  %A.i8ptr = bitcast float* %A to i8*
  %0 = call i8* @llvm_hpvm_ocl_argument_ptr(i8* %graph.SgemmLeaf, i8* %A.i8ptr, i32 0, i64 %bytesA, i1 true, i1 false)
  %bytesA.ptr = alloca i64
  store i64 %bytesA, i64* %bytesA.ptr
  %bytesA.i8ptr = bitcast i64* %bytesA.ptr to i8*
  call void @llvm_hpvm_ocl_argument_scalar(i8* %graph.SgemmLeaf, i8* %bytesA.i8ptr, i32 1, i64 ptrtoint (i64* getelementptr (i64* null, i32 1) to i64))
; Call modified allocation node function to compute the size of required
; local memory
  %3 = call %struct.out.Allocation @Allocation1(i32 %block_x, i32 %block_y)
  %4 = extractvalue %struct.out.Allocation %3, 1
  call void @llvm_hpvm_ocl_argument_shared(i8* %graph.SgemmLeaf, i32 12, i64 %4)
; ... set remaining kernel arguments

; Compute local and global workgroup sizes
  %LocalWGSize = alloca [2 x i64]
  %LocalWGSize.0 = bitcast [2 x i64]* %LocalWGSize to i64*
  %5 = sext i32 %block_x to i64
  store i64 %5, i64* %LocalWGSize.0
  %LocalWGSize.1 = getelementptr i64* %LocalWGSize.0, i64 1
; ...
  %GlobalWGSize = alloca [2 x i64]
; ...
; Runtime call to launch kernel
  %event.SgemmLeaf = call i8* @llvm_hpvm_ocl_executeNode(i8* %graph.SgemmLeaf, i32 2, i64* %LocalWGSize.0, i64* %GlobalWGSize.0)
; Runtime call to wait fo kernel completion
  call void @llvm_hpvm_ocl_wait(i8* %graph.SgemmLeaf)
; Free allocated memory.
; In our implementation, these are no-ops, since we free memory when
; no longer required as determined by the memory tracker.
  call void @llvm_hpvm_ocl_free(i8* %0)
  call void @llvm_hpvm_ocl_free(i8* %1)
  call void @llvm_hpvm_ocl_free(i8* %2)

  ret %emptyStruct.20 undef
}

; Function called by launch runtime call. Used to perform
; - argument unpacking
; - call to the genFunc that is generated by the GPU backend
define i8* @LaunchDataflowGraph(i8* %data.addr) {
entry:
; Argument unpacking
  %A.addr = bitcast i8* %data.addr to float**
  %A = load float** %A.addr
  %nextArg = getelementptr float** %A.addr, i64 1
  %bytesA.addr = bitcast float** %nextArg to i64*
  %bytesA = load i64* %bytesA.addr
; ... remaining argument unpacking

; Call x86 function that launches the GPU kernel
  %SgemmRoot2.output = call %emptyStruct.20 @SgemmRoot2(float* %A, i64 %bytesA, i32 %lda, float* %B, i64 %bytesB, i32 %ldb, float* %C, i64 %bytesC, i32 %ldc, i32 %k, float %alpha, float %beta, i32 %block_x, i32 %block_y, i32 %grid_x, i32 %grid_y)
  %SgemmRoot2.output.addr = bitcast i8* %data.addr to %emptyStruct.20*
  store %emptyStruct.20 %SgemmRoot2.output, %emptyStruct.20* %SgemmRoot2.output.addr
  ret i8* null
}
\end{lstlisting}

\listoftodos

%% D. References
\clearpage
\renewcommand\thesection {B}
\setcounter{page}{1}
% {\footnotesize{
\bibliographystyle{plain}
\bibliography{hetero,optimization}

\end{document}